\documentclass[a4paper,12pt]{article}%
\usepackage[affil-it]{authblk}
\usepackage[authoryear]{natbib}
\usepackage{bbm}
\usepackage{amsmath}
\usepackage{amssymb}
\usepackage{mathrsfs}
\usepackage{dsfont}
\usepackage{amsthm}
\usepackage{geometry}
\usepackage{threeparttable}
\usepackage{amsmath}
\usepackage{amssymb}
\usepackage{mathrsfs}
\usepackage{dsfont}
\usepackage{amsthm}
\usepackage{geometry}
\usepackage{color}
\usepackage{bbm}%
\usepackage{amsfonts}%
\usepackage{graphicx}
\providecommand{\U}[1]{\protect \rule{.1in}{.1in}}
\newtheorem{theorem}{Theorem}
\newtheorem{assumption}{Assumption}
\newtheorem{lemma}{Lemma}

\newtheorem{definition}{Definition}

\newcommand{\nc}{\newcommand}
\nc{\norm}{\mathcal{N}}
\nc{\E}{\mathcal{E}}
\nc{\bx}{{\bf x}}
\nc{\bX}{{\bf X}}
\nc{\by}{{\bf y}}
\nc{\IG}{\mathcal{IG}}
\nc{\dd}[2]{\frac{\partial #1}{\partial #2}}
\nc{\lhat}[1][i]{\hat \lambda_{#1}^{-1(g)}}
\nc{\what}[1][j]{\hat \omega_{#1}^{-1(g)}}
\nc{\bone}{{\bf 1}}
\nc{\Li}{\hat \Lambda^{-1(g)}}
\nc{\Oi}{\hat \Omega^{-1(g)}}
\nc{\dps}{\displaystyle}
\nc{\tr}{\text{tr}}

\def \E{\mathbb{E}}

\begin{document}

\title{A Simple and Efficient Estimation Method for Models with Nonignorable Missing
	Data  }
\author{Chunrong Ai\\Department of Economics, University of Florida\\
	tsinghua@ufl.edu
\and Oliver Linton\\Faculty of Economics, University of Cambridge \\
obl20@cam.ac.uk
\and Zheng Zhang\\Institute of Statistics and Big Data, Renmin University of China\\
zhengzhang@ruc.edu.cn} 
\maketitle

\begin{abstract}
This paper proposes a simple and efficient estimation procedure for the model with non-ignorable missing data studied by {\color{black}\cite{morikawa2016semiparametric}}. Their semiparametrically efficient estimator requires explicit nonparametric estimation and so suffers from the curse of dimensionality and requires a bandwidth selection. We propose an estimation method based on the Generalized Method of Moments (hereafter GMM). Our method is consistent and asymptotically normal regardless of the number of moments chosen. Furthermore, if the number of moments increases appropriately our estimator can achieve the semiparametric efficiency bound derived in {\color{black}\cite{morikawa2016semiparametric}}, but under weaker regularity conditions. Moreover, our proposed estimator and its consistent covariance matrix are easily computed with the widely available GMM package. We propose two data-based methods for selection of the number of moments. A small scale simulation study reveals that the proposed estimation indeed out-performs the existing alternatives in finite samples. 
\end{abstract}

	\section{Introduction}
	Missing data is common in many fields of applications. One way to deal with the missing data problem is to delete observations
	containing missing data. In doing so we may produce biased estimates and erroneous conclusions, depending on the data missing mechanism. If data are missing completely at random, standard estimation and inference procedures are still consistent when the missing data observations are ignored, see \cite{heitjan1996distinguishing}, \cite{little1988test} among others. If data are missing at random (MAR) in the sense that the propensity of missingness depends only on the observed covariates, consistent estimation can still be obtained through covariate balancing, see \cite{rubin1976comparing, rubin1976inference}, \cite{little1989analysis}, \cite{robins1995semiparametric}, \cite{robins1995analysis}, \cite{bang2005doubly}, \cite{qin2007empirical}, \cite{chen2008semiparametric}, \cite%
	{tan2010bounded}, \cite{rotnitzky2012improved}, \cite{little2014statistical}
	among others. In many applications, data are missing not at random (MNAR). For example, the income question in sample surveys is often not answered by people at the top end of the distribution, that is, their response frequency depends on an outcome variable that is often the key focus. An investigator is examining the effect of sleep on pain by calling subjects daily to ask them about last night's sleep and their pain today. Patients who are experiencing severe pain are more likely to not come to the phone leaving the data missing for that particular day; again this would violate the MAR assumption. From political science, roll-call votes, which measure legislatures ideological positions, are subject to non-ignorable nonresponse because, unsurprisingly, politicians behave strategically. 
	In the MNAR case, the parameter of interest may not even be identified (e.g., \cite{robins1997toward}), let
	alone be consistently estimated. To be more specific, let $T\in \{0,1\}$ denote the binary random variable indicating the missing status of the outcome variable $Y$: $Y$ is observed if $T$ takes the value one and $Y$ is not observed if $T$ takes the value zero. Let $\boldsymbol{X}$ denote a vector of explanatory variables, let $\pi (\boldsymbol{x},y)=\mathbb{P}(T=1|%
	\boldsymbol{X}=\boldsymbol{x},Y=y)$ denote the propensity score function and let $f_{Y|\boldsymbol{X}}(y|\boldsymbol{x})$ denote the conditional density function of $Y$ given $\boldsymbol{X}$. \cite{robins1997toward} shows that
	if both the propensity score function and the conditional density function are completely unknown, the joint \ distribution of $(T,Y)$ given $\boldsymbol{X}$ is
	not point identifiable. In this case, a necessary identification condition is the parameterization of either the propensity score function or the conditional density function. Molenberghs and Kenward (2007) proposes the parameterization of both the propensity score function and the conditional density function as an identification strategy, while \cite{sverchkov2008new} and \cite%
	{riddles2016propensity} parameterize the propensity score function and only a component of the conditional density function: $f_{Y|\boldsymbol{X},T}(y|\boldsymbol{x},T=1)$. \\

	If the joint distribution is not the parameter of interest, the identification strategy above can be modified. For example, if the parameter of interest is the conditional density of $Y$ given $\boldsymbol{X}$ (i.e., $f_{Y|\boldsymbol{X}}(y|\boldsymbol{x})$),
	parameterization of the propensity score function is not needed. However, parameterization of $f_{Y|\boldsymbol{X}}(y|\boldsymbol{x})$ in this case is not sufficient for identification due to missing data. \cite{tang2003analysis}
	suggests parameterization of the marginal density $f_{\boldsymbol{X}}(\boldsymbol{x})$ as well, while \cite{zhao2015semiparametric} imposes an exclusion
	retriction. In both studies, $f_{Y|\boldsymbol{X}}(y|\boldsymbol{x})$ is identified and consistently estimated. \\
	
	We consider estimation of the parameter $\theta _{0}=\mathbb{E}[U(\boldsymbol{X},Y)]$,
	where $U(\cdot )$ is any known function. We suppose that the propensity score $\pi$ is parameterized but do not restrict the conditional density function of the outcome variable. 
	In earlier work in this framework, either the coefficients in the propensity score
	function are known or consistently estimated from an external sample (\cite{kim2011semiparametric}) or an exclusion restriction is imposed (\cite{wang2014instrumental} and \cite{shao2016semiparametric}). \cite{morikawa2016semiparametric} study the
	efficient estimation of $\theta _{0}$. They derive the efficient score
	function (and hence the semiparametric efficiency bound) for $\theta _{0}$ in this model. They propose to estimate the
	efficient score function by estimating $f_{Y|\boldsymbol{X},T}(y|,%
	\boldsymbol{x},1)$ by a working parametric model (MK1) or by kernel nonparametric estimation (MK2). Their approach MK1 is not efficient unless the working parametric model is correct, although it is consistent.
	Their method MK2 suffers from the curse of dimensionality (their smoothness conditions depend on the dimensionality of the covariates through their conditions C14) and the bandwidth selection problem (about which they give no guidance). \\
	
	We study the same estimation problem as in \cite{morikawa2016semiparametric} but propose a simpler yet equally efficient estimation procedure. Our proposed method does not require explicit nonparametric estimation and hence does not suffer
	from the curse of dimensionality. The proposed
	estimator is motivated by the key insight that the model parameter
	satisfies a parametric conditional moment restriction, of which the
	semiparametric efficiency bound is identical to the bound derived in \cite%
	{morikawa2016semiparametric}. The conditional moment restriction is then turned
	into an expanding set of unconditional moment restrictions and the parameter
	of interest is estimated by applying the widely available and easy to
	compute GMM estimation (see Hansen (1982)). Under some sufficient conditions, we establish that the proposed
	estimator is consistent and asymptotically normally distributed even if the
	set of unconditional moment restrictions does not expand, thereby freeing us
	from the curse of dimensionality and the bandwidth selection problem; when the set does
	expand, the proposed estimator attains the semiparametric efficiency bound.
	This is in contrast with the MK2 method of \cite{morikawa2016semiparametric}%
	, which is inconsistent if the bandwidth does not go to zero at a certain rate.\\
	
	The paper is organized as follows. Section 2 describes the estimation,
	Section 3 derives the large sample properties of the estimator, Section 4 provides a consistent asymptotic variance estimator, Section 5 suggests two data driven approaches to determine the number of unconditional moment restrictions, Section 6 reports on a small scale simulation study, followed by some concluding remarks in Section 7. All technical proofs are relegated to the Appendix.
	
\section{Basic Framework and Estimation}
We begin by setting up the basic framework. Denote $\boldsymbol{Z}=(%
\boldsymbol{X}^{\top },Y)^{\top }$. The following assumption shall be maintained throughout the paper:

\ \\
\textbf{Assumption 2.1}. (i) Parameterization of data missing mechanism: $\mathbb{P}(T=1|Y,\boldsymbol{X})=\pi (Y,\boldsymbol{X};\gamma _{0})=\pi (%
\boldsymbol{Z};\gamma _{0})$ holds for some\ known function $\pi (.;.)$, where $\gamma _{0}\in \mathbb{R}^{p}$ for some known $p\in \mathbb{N}$ is the true (unknown) value; (ii) exclusion restriction: there exists some
nonresponse instrument variables $\boldsymbol{X}_{1}$ in $\boldsymbol{X}=(%
\boldsymbol{X}_{1}^{\top },\boldsymbol{X}_{2}^{\top })^{\top }$ so that $%
\boldsymbol{X}_{2}$ is independent of $T$ given both $\boldsymbol{X}_{1}$
and $Y$; and (iii) the parameter of interest is $\theta _{0}=\mathbb{E}[U(%
\boldsymbol{Z})]$ for some known function $U(\cdot )$.\\

Under Assumption 2.1 and by applying the law of iterated expectations, we obtain the following conditional moment restrictions:

\begin{align}
& \mathbb{E}\left[ 1-\frac{T}{\pi (\boldsymbol{Z};\gamma _{0})}\bigg|\boldsymbol{X%
}\right] =0,  \label{sequential1} \\
& \mathbb{E}\left[ \theta _{0}-\frac{T}{\pi (\boldsymbol{Z};\gamma _{0})}U(%
\boldsymbol{Z})\right] =0,  \label{sequential2}
\end{align}%
which will form the basis for the proposed estimation. We notice that the parameters of interest in (\ref{sequential1})-(\ref{sequential2}) are finite dimensional (and there is no explicit infinite dimensional nuisance parameter) and can be easily
estimated with GMM estimation. We also notice that it is a special case of
the model studied in \cite{ai2012semiparametric}. By applying their result (Remark
2.1, pp. 446), we obtain the semiparametric efficiency bound for model (\ref{sequential1})-(%
\ref{sequential2}), which is identical to the bound derived in \cite%
{morikawa2016semiparametric}, thereby suggesting a simple and efficient estimation.	\\

The (nuisance) parameter $\gamma _{0}$ is identified by (\ref{sequential1}) and the parameter of interest $\theta _{0}$ is identified by (\ref{sequential2}). The following condition shall also be maintained throughout the paper:	

\ \\	
\textbf{Assumption 2.2}. The parameter space $\Gamma $ is a compact subset
of $\mathbb{R}^{p}$. The true value $\gamma _{0}$ lies in the interior of $%
\Gamma $ and is the only solution to (\ref{sequential1}). {\color{black}The parameter space $\Theta$ is a compact subset of $\mathbb{R}$ and the true value $\theta_0$ lies in the interior of $\Theta$.}	\\

To estimate model (\ref{sequential1})-(\ref{sequential2}), we first turn it
into a set of unconditional moment restrictions. We work with a set of known basis functions: for each integer $\ K\in 
\mathbb{N}\ \text{with}\ K\geq p\ $, let
\begin{equation*}
u_{K}(\boldsymbol{X})=(u_{1K}(\boldsymbol{X}),\ldots ,u_{KK}(\boldsymbol{X}%
))^{\top }.\ 
\end{equation*}
{\color{black}Discussion on the choice of $u_K(\boldsymbol{X})$ and its properties can be found in Section \ref{sec:uK} in Appendix}. Model (\ref{sequential1})-(\ref%
{sequential2}) {\color{black}implies the unconditional moment restrictions}:

\begin{align}
& \mathbb{E}\left[ \left( 1-\frac{T}{\pi (\boldsymbol{Z};\gamma _{0})}%
\right) u_{K}(\boldsymbol{X})\right] =0,  \label{uncond1} \\
& \mathbb{E}\left[ \theta _{0}-\frac{T}{\pi (\boldsymbol{Z};\gamma _{0})}U(%
\boldsymbol{Z})\right] =0.  \label{uncond2}
\end{align}%
To avoid redundant moment restrictions, we require $\mathbb{E}\left[ u_{K}(%
\boldsymbol{X})u_{K}(\boldsymbol{X})^{\top }\right] $ to be nonsingular for
every $K$. The following somewhat stronger identification condition shall be maintained throughout the paper:

\bigskip

\textbf{Assumption 2.2'}. The parameter space $\Gamma $ is a compact subset
of $\mathbb{R}^{p}$. The true value $\gamma _{0}$ lies in the interior of $%
\Gamma $ and is the only solution to (\ref{uncond1}). {\color{black}The parameter space $\Theta$ is a compact subset of $\mathbb{R}$ and the true value $\theta_0$ lies in the interior of $\Theta$.}\\

We can estimate the parameter of interest by the GMM
method. Let $\{T_{i},\boldsymbol{Z}_{i}\}_{i=1}^{N}$ denote an $i.i.d%
\text{.}$ sample drawn from the joint distribution of $(T,\boldsymbol{Z})$.
Denote

\begin{eqnarray*}
	\boldsymbol{G}_{K}(\gamma ,\theta ) :&=&\left( \sum_{i=1}^{N}\left[ 1-\frac{%
		T_{i}}{\pi (\boldsymbol{Z}_{i};\gamma )}\right] u_{K}(\boldsymbol{X}%
	_{i})^{\top },\sum_{i=1}^{N}\left[ \theta -\frac{T_{i}}{\pi (\boldsymbol{Z}%
		_{i};\gamma )}U(\boldsymbol{Z}_{i})\right] \right) ^{\top } \\
	&=&\sum_{i=1}^{N}g_{K}(T_{i},\boldsymbol{Z}_{i};\gamma ,\theta )\text{,}
\end{eqnarray*}
{\color{black} where $g_{K}(T,\boldsymbol{Z};\gamma ,\theta ):=\left( \left[ 1-\frac{%
		T}{\pi (\boldsymbol{Z};\gamma )}\right] u_{K}(\boldsymbol{X})^{\top },   \theta -\frac{T}{\pi (\boldsymbol{Z};\gamma )}U(\boldsymbol{Z})\right) ^{\top }$.} The GMM estimator of $\gamma _{0}$ and $\theta _{0}$ is defined as 
\begin{equation*}
(\check{\gamma},\check{\theta})=\arg \underset{\gamma \in \Gamma ,\theta\in\Theta }{%
	\min }\boldsymbol{G}_{K}(\gamma ,\theta )^{T}\cdot \mathbf{W}\cdot 
\boldsymbol{G}_{K}(\gamma ,\theta )
\end{equation*}%
where $\mathbf{W}$ is a $(K+1)\times (K+1)$ symmetric weighting matrix. For every fixed $K\geq p$, \cite{hansen1982large} shows that, under some regularity conditions, the estimator
\begin{equation}
{(\check{\gamma}-\gamma _{0},\check{\theta}-\theta _{0})=\color{black}O_{p}({N}^{-1/2})}
\label{Hansen1}
\end{equation}%
{\ is asymptotically normally distributed, but generally not the best unless
	the best weighting matrix is used. The best weighting matrix is the inverse
	of }%
\begin{equation*}
\boldsymbol{D}_{(K+1)\times (K+1)}:=\mathbb{E}\left[ g_{K}(T,\boldsymbol{Z};\gamma
_{0},\theta _{0})g_{K}(T,\boldsymbol{Z};\gamma _{0},\theta _{0})^{\top }%
\right] .
\end{equation*}%
The best estimator (within the class defined by the specific unconditional moments) is defined as 
\begin{equation*}
(\overline{\gamma },\overline{\theta })=\arg \underset{\gamma \in \Gamma
	,\theta\in\Theta }{\min }\boldsymbol{G}_{K}(\gamma ,\theta )^{T}\cdot \boldsymbol{D}%
_{(K+1)\times (K+1)}^{-1}\cdot \boldsymbol{G}_{K}(\gamma ,\theta ).
\end{equation*}%
Suppose that the propensity score function is differentiable with respect to 
$\gamma $. Denote 
\begin{align*}
\boldsymbol{B}_{(K+1)\times (p+1)} =\nabla _{\gamma ,\theta }\mathbb{E}%
\left[ \frac{1}{N}\boldsymbol{G}_{K}(\gamma _{0},\theta _{0})\right] =\left( 
\begin{array}{cc}
\mathbb{E}\left[ u_{K}(\boldsymbol{X})\frac{\nabla _{\gamma }\pi (%
	\boldsymbol{Z};\gamma _{0})^{\top}}{\pi (\boldsymbol{Z};\gamma _{0})}\right] 
\text{,}\vspace{4mm} & \mathbf{0}_{K\times 1} \\ 
\mathbb{E}\left[ U(\boldsymbol{Z})\frac{\nabla _{\gamma }\pi (\boldsymbol{Z}%
	;\gamma _{0})^{\top}}{\pi (\boldsymbol{Z};\gamma _{0})}\right] \text{,} & 1
\end{array}
\right)
\end{align*}
and
\begin{align*} \boldsymbol{V}_{K} =\left\{ \left( \boldsymbol{B}_{(K+1)\times
	(p+1)}\right) ^{\top }\boldsymbol{D}_{(K+1)\times (K+1)}^{-1}\left( 
\boldsymbol{B}_{(K+1)\times (p+1)}\right) \right\} ^{-1}.
\end{align*}
Hansen (1982) shows that, for every fixed $K\geq p$, 
\begin{equation}
\boldsymbol{V}_{K}^{-1/2}\binom{\sqrt{N}(\overline{\gamma }-\gamma _{0})}{%
	\sqrt{N}(\overline{\theta }-\theta _{0})}\rightarrow N\left(
0,I_{(p+1)\times (p+1)}\right) \text{ in distribution.}  \label{Hansen2}
\end{equation}
Since the best weighting matrix depends on the unknown parameter value, the
best estimator $(\overline{\gamma },\overline{\theta })$ is infeasible.
Hansen (1982) suggests the following two-step procedure:\\

\qquad Step I. Compute the initial $\sqrt{N}$-consistent estimator 
{\color{black}\begin{align*}
	&\widehat{\boldsymbol{W}}_0:=\begin{pmatrix}
	\frac{1}{N}\sum_{i=1}^Nu_K(\bold{X}_i)u_K(\bold{X}_i)^{\top} & \boldsymbol{0}_{K\times 1} \\[2mm]
	\boldsymbol{0}_{K\times 1}^{\top} & 1
	\end{pmatrix}\ ,\\
	&(\check{\gamma},\check{\theta})=\arg \underset{(\gamma,\theta) \in \Gamma \times \Theta }{%
		\min }\boldsymbol{G}_{K}(\gamma ,\theta )^{T}\cdot \widehat{\boldsymbol{W}}^{-1}_0\cdot \boldsymbol{G}_{K}(\gamma
	,\theta ).
	\end{align*}}
\qquad \qquad\ \qquad

\qquad Step II. Compute the best weighting matrix and the best estimator {\ 
	\begin{equation*}
	\hat{\boldsymbol{D}}_{(K+1)\times (K+1)}:=\frac{1}{N}\sum_{i=1}^{N}g_{K}(T_{i},%
	\boldsymbol{Z}_{i};\check{\gamma},\check{\theta})g_{K}(T_{i},\boldsymbol{Z}%
	_{i};\check{\gamma},\check{\theta})^{\top }\;\text{,}
	\end{equation*}%
} 
\begin{equation*}
(\widehat{\gamma },\widehat{\theta })=\arg \underset{\gamma \in \Gamma
	,\theta \in \Theta }{\min }\boldsymbol{G}_{K}(\gamma ,\theta )^{T}\cdot 
\widehat{\boldsymbol{D}}_{(K+1)\times (K+1)}^{-1}\cdot \boldsymbol{G}%
_{K}(\gamma ,\theta ).
\end{equation*} 

Hansen (1982) establishes that, for {every fixed $K\geq p$}, 
\begin{equation}
\boldsymbol{V}_{K}^{-1/2}\binom{\sqrt{N}(\widehat{\gamma }-\gamma _{0})}{%
	\sqrt{N}(\widehat{\theta }-\theta _{0})}\rightarrow N\left( 0,I_{(p+1)\times
	(p+1)}\right) \text{ in distribution.}  \label{Hansen3}
\end{equation}%
Moreover, denote 
\begin{align*}
\widehat{\boldsymbol{B}}_{(K+1)\times (p+1)} :=\left( 
\begin{array}{cc}
N^{-1}\sum_{i=1}^{N}u_{K}(\boldsymbol{X}_{i})\frac{\nabla _{\gamma }\pi (%
	\boldsymbol{Z}_{i};\widehat{\gamma })^{\top}}{\pi (\boldsymbol{Z}_{i};\widehat{%
		\gamma })}\text{,}\vspace{4mm} & \mathbf{0}_{K\times 1} \\ 
N^{-1}\sum_{i=1}^{N}U(\boldsymbol{Z}_{i})\frac{\nabla _{\gamma }\pi (%
	\boldsymbol{Z}_{i};\widehat{\gamma })^{\top}}{\pi (\boldsymbol{Z}_{i};\widehat{%
		\gamma })}\text{,} & 1%
\end{array}%
\right)
\end{align*}
and
\begin{align*}
\widehat{\boldsymbol{V}}_{K} :=\left\{ \left( \widehat{%
	\boldsymbol{B}}_{(K+1)\times (p+1)}\right) ^{\top }\widehat{\boldsymbol{D}}%
_{(K+1)\times (K+1)}^{-1}\left( \widehat{\boldsymbol{B}}_{(K+1)\times
	(p+1)}\right) \right\} ^{-1}.
\end{align*}%
Hansen (1982) proves that, for {every fixed $K\geq p$}, 
\begin{equation}
\widehat{\boldsymbol{V}}_{K}\rightarrow \boldsymbol{V}_{K}\text{ in
	probability.}  \label{Hansen4}
\end{equation}

The best estimator (within the class defined by the specific unconditional moments) is generally not semiparametrically efficient. To obtain the efficient
estimator, we shall allow $K$ to increase with the sample size at the rate  {\color{black}$
	o(N^{1/3})$} so that $\{u_{K}(\boldsymbol{X})\}$ span the space of measureable
functions (see also \cite{geman1982nonparametric} and \cite%
{newey1997convergence}). In the next two sections, we shall establish that results in (\ref{Hansen1})-(\ref{Hansen4}) still hold with expanding {\color{black}$K=o(N^{1/3})$}. \\

The advantage of our proposed estimator over the existing estimators is evident. Our
estimation problem is a parametric one, requiring no modeling of or
nonparametric estimation of $f_{Y|\boldsymbol{X},T}(y|x,1)$. In contrast,
the estimators proposed in \cite{riddles2016propensity} and \cite%
{morikawa2016semiparametric} could be inconsistent if $f_{Y|\boldsymbol{X}%
	,T}(y|x,1)$ is incorrectly specified or suffers from the curve of
dimensionality and bandwith selection problem of the nonparametric
estimation of $f_{Y|\boldsymbol{X},T}(y|x,1)$.

\section{Asymptotic Theory}
In this section, we show that results in {\ (\ref{Hansen1})- (%
	\ref{Hansen3}) still hold with expanding }$K$, all technical  proof can be found in the supplemental material \cite{alz2018}. First, we establish the convergence rate of the first step estimator $(\check{\gamma}%
,\check{\theta})$. 

\begin{theorem}
	Under Assumptions 2.1-2.2 and Assumptions \ref{as:id}, \ref{as:suppX}, \ref{as:eigen}, \ref{as:iid}, \ref{as:pi}, and \ref{as:K&N} listed in Appendix,
	with $K=o(N^{1/3})$, the first step estimator satisfies 
	\begin{equation*}
	(\check{\gamma}-\gamma _{0},\check{\theta}-\theta _{0})=O_{p}\left(N^{-1/2}\right).
	\end{equation*}
\end{theorem}

Next, we establish the large sample properties of the infeasible
best estimator $(\overline{\gamma },\overline{\theta })$ without imposing
the smoothness Assumptions \ref{as:proj_smooth} and \ref{as:boundS} listed in  Appendix. 

\begin{theorem}
	Under Assumptions 2.1-2.2 and Assumptions \ref{as:id}, \ref{as:suppX}, \ref{as:eigen}, \ref{as:iid}, \ref{as:pi}, and \ref{as:K&N} listed in
	Appendix, with $K=o(N^{1/3})$, the infeasible best estimator satisfies 
	\begin{equation*}
	\boldsymbol{V}_{K}^{-1/2}\binom{\sqrt{N}(\overline{\gamma }-\gamma _{0})}{%
		\sqrt{N}(\overline{\theta }-\theta _{0})}\rightarrow N\left(
	0,I_{(p+1)\times (p+1)}\right) \text{ in distribution.}
	\end{equation*}
\end{theorem}

{If in addition the smoothness Assumptions \ref{as:proj_smooth} and \ref%
	{as:boundS} are
	satisfied, the next result shows that $\boldsymbol{V}_{K}\rightarrow 
	\boldsymbol{V}_{eff}$} in probability, where $\boldsymbol{V}_{eff}$ is the
semiparametric efficiency bound of $(\gamma _{0},\theta _{0})$ {\color{black}derived in \cite{morikawa2016semiparametric},  see Lemma 1 in Section 8.3 of Appendix.} 

\begin{theorem}
	Under Assumption 2.1-2.2 and Assumption 1-\ref{as:K&N} listed in Appendix, with $%
	K=o(N^{1/3})$, we obtain 
	\begin{equation*}
	\boldsymbol{V}_{K}\rightarrow \boldsymbol{V}_{eff}\text{ in probability.}
	\end{equation*}
\end{theorem}

By Theorem 1-3, the infeasible best estimator attains the semiparametric efficiency bound. The next result establishes
the equivalence between the best estimator $(\widehat{\gamma },\widehat{%
	\theta })$ and the infeasible best estimator $(\overline{\gamma },\overline{%
	\theta })$, implying that the best estimator also attains the semiparametric
efficiency bound. 

\begin{theorem}
	Under Assumption 2.1-2.2 and Assumption 1-\ref{as:K&N} listed in Appendix, with $%
	K=o(N^{1/3})$, we obtain 
	\begin{equation*}
	\binom{\sqrt{N}(\overline{\gamma }-\widehat{\gamma })}{\sqrt{N}(\overline{%
			\theta }-\widehat{\theta })}=o_{p}(1).
	\end{equation*}
\end{theorem}

\section{Variance Estimation}
In order to conduct statistical inference, we need a consistent covariance
estimator. Notice that {\ (\ref{Hansen1}) } implies that $\widehat{\boldsymbol{V}}_{K}
$ is a consistent estimator of $\boldsymbol{V}_{K}$ for every fixed $K\geq p$%
. We now show that this result still holds true with expanding $K$, thereby
providing a consistent covariance estimator. 

\begin{theorem}
	Under Assumption 2.1-2.2 and Assumption 1-\ref{as:K&N} listed in Appendix, with $%
	K=o(N^{1/3})$, we obtain%
	\begin{equation*}
	\widehat{\boldsymbol{V}}_{K}\rightarrow \boldsymbol{V}_{K}\text{ in
		probability.}
	\end{equation*}
\end{theorem}

We notice that our covariance estimator is much simpler and more natural
than the one suggested in \cite{morikawa2016semiparametric}, which requires
nonprametric estimation of  $f_{Y|\boldsymbol{X},T}(y|x,1)$ and tends to have
poor performance in finite samples. Our covariance estimator is the GMM
covariance estimator and is easily computed by existing statistical
packages.  

\section{Selection of $K$}
The large sample properties of the proposed estimator established in the
previous sections allow for a wide range of values for $K$, and theoretically the sensitivity of the estimator
to the choice of $K$ is not so pronounced, it affects higher order terms in a way
that does not affect consistency and asymptotic normality. Nevertheless, there may be some 
higher order effect of the choice of $K$ on perfomance. In this section, we present two data-driven approaches to
select $K$. Both approaches strike a balance between bias and variance.  \\

\textbf{Covariate balancing approach}. The first approach attempts to
balance the distribution of the covariates between the whole population and
the non-missing population through weighting. Notice that 
\begin{equation*}
\mathbb{E}\left[ \frac{T}{\pi (\boldsymbol{Z};\gamma _{0})}I(X_{j}\leq x_{j})%
\right] =\mathbb{E}[I(X_{j}\leq x_{j})]\ ,\ j\in \{1,...,r\}\ ,
\end{equation*}%
where $X_{j}$ is the $j^{th}$ component of $\boldsymbol{X}$ and $I(X_{j}\leq
x_{j})$ is the indicator function. Obviously the propensity score function $\pi (%
\boldsymbol{Z};\gamma _{0})$ plays the role of balancing. Notice that the
estimator $\hat{\gamma}$ depends on $K.$ For a given $K$, we compute   
\begin{equation*}
\hat{F}_{N,K}^{j}(x_{j}):=\frac{1}{N}\sum_{i=1}^{N}\frac{T_{i}}{\pi (%
	\boldsymbol{X}_{i};\hat{\gamma})}I(X_{ij}\leq x_{j}),\;j\in \{1,\ldots ,r\}%
\text{.}
\end{equation*}%
We compute the empirical distributions of the covariates 
\begin{equation*}
\tilde{F}_{N}^{j}(x_{j}):=\frac{1}{N}\sum_{i=1}^{N}I(X_{ij}\leq
x_{j}),\;j\in \{1,\ldots ,r\}\text{.}
\end{equation*}%
We choose the lowest $K$ so that the difference between $\{\hat{F}%
_{N,K}^{j}\}_{j=1}^{r}$ and $\{\hat{F}_{N}^{j}\}_{j=1}^{r}$ is small. Denote
the upper bound of $K$ by $\bar{K}$ (e.g. $\bar{K}=7$ in our simulation
studies). We choose $K\in \{1,...,\bar{K}\}$ to minimize the aggregate
Kolmogorov-Smirnov distance between $\{\hat{F}_{N,K}^{j}\}_{j=1}^{r}$ and $\{%
\hat{F}_{N}^{j}\}_{j=1}^{r}$: 
\begin{equation*}
\hat{K}=\arg \min_{K\in \{1,...,\bar{K}\}}{D}_{N}(K)=\sum_{j=1}^{r}%
\sup_{x_{j}\in \mathbb{R}}\left\vert \tilde{F}_{N}^{j}(x_{j})-\hat{F}%
_{N,K}^{j}(x_{j})\right\vert .
\end{equation*}

{\color{black}	
	\textbf{Higher order MSE approach}.  The second approach chooses $K$ to minimize the mean-squared error of the estimator. \cite{donald2009choosing}  derives the 
	higher-order asymptotic mean-square error (MSE) of a linear combination $\mathbf{t}^{\top } \hat{\gamma }$ for some fixed $\mathbf{t} \in \mathbb{R}^{p}$. \\
	
	Let $\check{\gamma }$ be some preliminary estimator. Define:
	\begin{align*} & \widehat{\Pi } (K ;\boldsymbol{t}) =\sum _{i =1}^{N}\hat{\xi }_{i i} \rho  (T_{i} ,\boldsymbol{X}_{i} ,Y_{i} ;\check{\gamma }) \cdot (\mathbf{t}^{\top } \hat{\Omega }_{p \times p}^{ -1} \tilde{\mathbf{\eta }}_{i})\text{,} \\
	& \hat{\Phi } (K ;\boldsymbol{t}) =\sum _{i =1}^{N}\hat{\xi }_{i i} \left \{\mathbf{t}^{\top } \hat{\Omega }_{p \times p}^{ -1} \left [\widehat{\mathbf{D}}_{i}^{ \ast } \rho  (T_{i} ,\boldsymbol{X}_{i} ,Y_{i} ;\check{\gamma })^{2} - \nabla _{\gamma }\rho  (T_{i} ,\boldsymbol{X}_{i} ,Y_{i} ;\check{\gamma })\right ]\right \}^{2} \\
	& \text{\quad \quad \quad \quad \quad \quad } -\mathbf{t}^{\top } \hat{\Omega }_{p \times p}^{ -1} (\hat{\Gamma }_{K \times p})^{\top } \hat{\Upsilon }_{K \times K}^{ -1} \hat{\Gamma }_{K \times p} \hat{\Omega }_{p \times p}^{ -1} \boldsymbol{t}\;\text{.}\end{align*}
	where $\rho(T_i,\boldsymbol{X}_i,Y_i;\check{\gamma})$, $\hat{\Omega}_{p\times p}$, $\tilde{\eta}_i$, $\hat{\xi}_{ii}$, $\hat{\boldsymbol{D}}_i^*$, $\hat{\Gamma}_{K\times p}$, and $\hat{\Upsilon}_{K\times K}$ are defined in Section 8.2 of Appendix.  Notice that $\widehat{\Pi}(K;\boldsymbol{t})^2/N$ is an estimate of the squared bias term derived in \cite{newey2004higher} and $\hat{\Phi } (K;\boldsymbol{t})$  is the asymptotic variance. \\
	
	The second approach chooses $K$ to minimize the following higher-order MSEs of $\hat{\gamma }_{j} ,j =1 ,\ldots  ,p$:
	\begin{align}S_{G M M} (K) =\sum _{j =1}^{p}\left \{\frac{1}{N} \widehat{\Pi } (K ;e_{j})^{2} +\hat{\Phi } (K ;e_{j})\right \}\;\text{,} \label{eq:criteria_K}\end{align}where $e_{j}$ is the $j^{t h}$ column of the $p$-dimensional identity matrix. In practice, we set the upper bound $\bar{K}$ and then choose $K \in \{1 ,2 ,\ldots  ,\bar{K}\}$ to minimize the criteria  \eqref{eq:criteria_K}}
.

\begin{table}
	\centering
	\caption{Simulation results under Scenorio I}
	\begin{threeparttable}
		{\fontsize{12pt}{15pt} \selectfont 
			\begin{tabular}{c|ccccccccccc}
				\hline \hline  
				\multicolumn{10}{c}{$n = 200$} \\
				\hline \\
				& $\hat{\alpha}$& $\hat{\beta}$ & $\hat{\theta}$ & $\hat{\alpha}_{MK}$ & $\hat{\beta}_{MK}$ & $\hat{\theta}_{MK}$ & $\tilde{\alpha}_{MAR}$ & $\tilde{\beta}_{MAR}$ & $\tilde{\theta}_{MAR}$ \\[1mm]		
				\hline\\ 
				Bias  &   0.028     &    -0.125   &   0.039   & 0.055     &  0.120   &  0.106  &  -0.997   &  0.167   & 0.301   \\
				Stdev  &  0.254    &  0.413   &  0.129   &  0.229   & 0.272  & 0.118  & 0.197   &0.266    & 0.101 \\
				MSE  &  0.065    & 0.186    & 0.018   &  0.055    & 0.088  &  0.025  &  1.033  & 0.099  & 0.101  \\
				CP  & --- & ---  &  0.908    & --- & ---  &   0.908    &--- &--- &  0.22   \\
				\hline \hline  
				\multicolumn{10}{c}{$n = 500$} \\
				\hline \\
				& $\hat{\alpha}$& $\hat{\beta}$ & $\hat{\theta}$ & $\hat{\alpha}_{MK}$ & $\hat{\beta}_{MK}$ & $\hat{\theta}_{MK}$ & $\tilde{\alpha}_{MAR}$ & $\tilde{\beta}_{MAR}$ & $\tilde{\theta}_{MAR}$ \\[1mm]		
				\hline\\ 
				Bias  &    0.011  &  -0.067   & 0.016   &  0.048   & 0.058  &  0.063  &   -0.966 & 0.220  & 0.299  \\
				Stdev  & 0.161     &  0.282   & 0.090   &  0.151   & 0.193  &  0.077  & 0.126 & 0.160  & 0.063 \\
				MSE  &  0.026    &  0.084   &  0.008  &  0.025   &  0.040 & 0.010   & 0.949  &  0.074 &0.093  \\
				CP  & --- & ---  &   0.928  & --- & ---  &  0.892    &--- &--- &  0.034 \\
				\hline \hline \multicolumn{10}{c}{}	
				\\[1mm]
				\hline \hline
				\multicolumn{10}{c}{$n = 1000$} 	\\  \hline \\
				& $\hat{\alpha}$& $\hat{\beta}$ & $\hat{\theta}$ & $\hat{\alpha}_{MK}$ & $\hat{\beta}_{MK}$ & $\hat{\theta}_{MK}$ & $\tilde{\alpha}_{MAR}$ & $\tilde{\beta}_{MAR}$ & $\tilde{\theta}_{MAR}$\\[1mm]
				\hline\\
				Bias &   0.005   &  -0.040   & 0.008   &  0.034   &  0.023  &  0.040  & -0.962  &  0.235 & 0.298\\
				Stdev &  0.103    &  0.187   & 0.065   &  0.102   & 0.132  & 0.055   & 0.078 & 0.099  & 0.045 \\
				MSE  &  0.010    &  0.036   &  0.004  &  0.011   & 0.018  &   0.004 &0.932  &0.065   &0.091\\
				CP  & --- & ---  &   0.934  & --- & ---  &   0.906    & --- & --- &0.012  \\	
				\hline	\hline
			\end{tabular}}
			{\fontsize{9.5pt}{12pt} \selectfont Stdev: standard deviation ; MSE: mean squared error;
				CP: coverage probability. The bandwith used in computing the nonparametric kernel estimators $(\hat{\alpha}_{MK},\hat{\beta}_{MK},\hat{\theta}_{MK})$ is $h=0.15$.	 }
		\end{threeparttable}
	\end{table}
	
	\begin{table}
		\centering
		\caption{Simulation results under Scenorio II}
		\begin{threeparttable}
			{\fontsize{12pt}{15pt} \selectfont 
				\begin{tabular}{c|ccccccccccc}
					\hline \hline  
					\multicolumn{10}{c}{$n = 200$} \\
					\hline \\
					& $\hat{\alpha}$& $\hat{\beta}$ & $\hat{\theta}$ & $\hat{\alpha}_{MK}$ & $\hat{\beta}_{MK}$ & $\hat{\theta}_{MK}$ & $\tilde{\alpha}_{MAR}$ & $\tilde{\beta}_{MAR}$ & $\tilde{\theta}_{MAR}$ \\[1mm]		
					\hline\\ 
					Bias  &   -0.208     &  0.096     & 0.084     &  -0.552    &   0.588  & 0.173   &  -2.053   &  1.215   &  0.530  \\
					Stdev  & 0.646     &  0.555   &  0.201   & 0.372    & 0.245  & 0.125  & 0.809   & 0.148   & 0.205 \\
					MSE  &  0.462    &  0.318   &  0.047  &  0.443    & 0.406  &  0.045  & 4.873   & 1.498  &0.323   \\
					CP  & --- & ---  &  0.95    & --- & ---  &   0.784    &--- &--- & 0.138   \\
					\hline \hline  
					\multicolumn{10}{c}{$n = 500$} \\
					\hline \\
					& $\hat{\alpha}$& $\hat{\beta}$ & $\hat{\theta}$ & $\hat{\alpha}_{MK}$ & $\hat{\beta}_{MK}$ & $\hat{\theta}_{MK}$ & $\tilde{\alpha}_{MAR}$ & $\tilde{\beta}_{MAR}$ & $\tilde{\theta}_{MAR}$ \\[1mm]		
					\hline\\ 
					Bias &  -0.081   &   0.040  & 0.044   & -0.313    &   0.392  &  0.122   & -1.924& 1.203 & 0.583 \\
					Stdev  & 0.406  &  0.363    & 0.131  &  0.261  &  0.186   &  0.085  &  0.175  &  0.064  & 0.132   \\
					MSE  &  0.171 & 0.134   & 0.019  &   0.166   &  0.188    & 0.022  &3.732 & 1.451& 0.357 \\
					CP  & --- & ---  &  0.932   & --- & ---  & 0.764     &--- &--- & 0.06  \\
					\hline \hline \multicolumn{10}{c}{}	
					\\[1mm]
					\hline \hline
					\multicolumn{10}{c}{$n = 1000$} 	\\  \hline \\
					& $\hat{\alpha}$& $\hat{\beta}$ & $\hat{\theta}$ & $\hat{\alpha}_{MK}$ & $\hat{\beta}_{MK}$ & $\hat{\theta}_{MK}$ & $\tilde{\alpha}_{MAR}$ & $\tilde{\beta}_{MAR}$ & $\tilde{\theta}_{MAR}$\\[1mm]
					\hline\\
					Bias & -0.036    & 0.019    & 0.019  &  -0.198   & 0.268     &   0.085   &-1.900 & 1.201  &  0.590\\
					Stdev  &  0.260  &    0.225    &  0.086 &  0.203    &   0.164 &  0.061  &  0.086  & 0.044 &  0.078  \\
					MSE  & 0.069  &  0.051   & 0.007  &  0.080   &  0.098    &  0.011    & 3.618 &  1.445 & 0.354 \\
					CP  & --- & ---  &  0.932   & --- & ---  &   0.768    & --- & --- & 0.018 \\	
					\hline	\hline
				\end{tabular}}
				{\fontsize{9.5pt}{14pt} \selectfont  Stdev: standard deviation ; MSE: mean squared error;
					CP: coverage probability. The bandwith used in computing the nonparametric kernel estimators $(\hat{\alpha}_{MK},\hat{\beta}_{MK},\hat{\theta}_{MK})$ is $h=0.05$.	
				}
			\end{threeparttable}
		\end{table}

		\begin{table}
			\caption{Simulation results under Scenorio III}
			\centering
			\begin{threeparttable}
				{\fontsize{12pt}{15pt} \selectfont 
					\begin{tabular}{c|ccccccccccc}
						\hline \hline  
						\multicolumn{10}{c}{$n = 200$} \\
						\hline \\
						& $\hat{\alpha}$& $\hat{\beta}$ & $\hat{\theta}$ & $\hat{\alpha}_{MK}$ & $\hat{\beta}_{MK}$ & $\hat{\theta}_{MK}$ & $\tilde{\alpha}_{MAR}$ & $\tilde{\beta}_{MAR}$ & $\tilde{\theta}_{MAR}$ \\[1mm]		
						\hline\\ 
						Bias  & 0.155       &  -0.171     &  0.003    &  0.047    &  0.015   &  0.071  &  -2.794   &  0.954   & -1.146   \\
						Stdev  &  0.584    &   0.585  &  0.155   & 0.376    &  0.190 & 0.131  &  1.395  & 0.396   & 0.263 \\
						MSE  &  0.365    &   0.372  & 0.024   &  0.144    & 0.036  &  0.022  & 9.758   & 1.069  & 1.384  \\
						CP  & --- & ---  &   0.934   & --- & ---  &   0.884   &--- &--- &   0.032  \\
						\hline \hline  	
						\multicolumn{10}{c}{$n = 500$} 	\\
						\hline \\	
						& $\hat{\alpha}$& $\hat{\beta}$ & $\hat{\theta}$ & $\hat{\alpha}_{MK}$ & $\hat{\beta}_{MK}$ & $\hat{\theta}_{MK}$ & $\tilde{\alpha}_{MAR}$ & $\tilde{\beta}_{MAR}$ & $\tilde{\theta}_{MAR}$ \\[1mm]		
						\hline\\ 
						Bias &  0.034   &   -0.036    &  0.000   &  0.012   &  0.012 &  0.034  & 0.782  & 0.355  &  0.123  \\
						Stdev  & 0.305    &  0.224  & 0.103   & 0.250   &  0.128  & 0.085  & 0.433   & 0.113  & 0.101    \\
						MSE  & 0.094  &  0.051   &  0.010  & 0.062  & 0.016   &   0.008   & 0.799  & 0.139 & 0.025  \\
						CP  & --- & ---  & 0.902  & --- & ---  & 0.894  & ---& --- &   0.698  \\	
						\hline \hline  \multicolumn{10}{c}{}	
						\\[1mm]
						\hline \hline
						\multicolumn{10}{c}{$n = 1000$}	\\ \hline\\
						& $\hat{\alpha}$& $\hat{\beta}$ & $\hat{\theta}$ & $\hat{\alpha}_{MK}$ & $\hat{\beta}_{MK}$ & $\hat{\theta}_{MK}$ & $\tilde{\alpha}_{MAR}$ & $\tilde{\beta}_{MAR}$ & $\tilde{\theta}_{MAR}$ \\[1mm]		
						\hline\\ 		
						Bias &  0.009    &  -0.010    & 0.002  &  0.002 & 0.009  &  0.017 &  0.728   & 0.372 &  0.126 \\
						Stdev  &  0.215  &   0.157   &  0.069  & 0.167   &  0.083   &   0.056 & 0.302  &0.078    &0.067 \\
						MSE  & 0.046  & 0.024   & 0.004  & 0.028   & 0.007    &  0.003  & 0.621  & 0.144  & 0.020 \\
						CP  & --- & ---  &   0.932   & --- & ---  &  0.934   & --- &--- &  0.454 \\	
						\hline \hline		
					\end{tabular}}
					{\fontsize{9.5pt}{14pt} \selectfont  Stdev: standard deviation ; MSE: mean squared error;
						CP: coverage probability. The bandwith used in computing the nonparametric kernel estimators $(\hat{\alpha}_{MK},\hat{\beta}_{MK},\hat{\theta}_{MK})$ is $h=0.1$.	
					}
				\end{threeparttable}
			\end{table}

			\begin{table}
				\caption{Simulation results under Scenorio IV}
				\centering
				\begin{threeparttable}
					{\fontsize{12pt}{15pt} \selectfont 
						\begin{tabular}{c|ccccccccccc}
							\hline \hline  
							\multicolumn{10}{c}{$n = 200$} \\
							\hline \\
							& $\hat{\alpha}$& $\hat{\beta}$ & $\hat{\theta}$ & $\hat{\alpha}_{MK}$ & $\hat{\beta}_{MK}$ & $\hat{\theta}_{MK}$ & $\tilde{\alpha}_{MAR}$ & $\tilde{\beta}_{MAR}$ & $\tilde{\theta}_{MAR}$ \\[1mm]		
							\hline\\ 
							Bias  &    0.097     &    -0.114   &   0.005   & -0.018     &  0.027   & 0.043   &  -1.002   & 1.003    &  0.136  \\
							Stdev  &  1.140    &  0.721   &  0.118   &  0.308   & 0.185  & 0.103  & 0.081   & 0.139   & 0.348 \\
							MSE  &  1.310    &  0.533   & 0.014   &  0.095    & 0.035  &  0.013  &  1.011  &  1.026 &  0.139 \\
							CP  & --- & ---  &   0.914   & --- & ---  &  0.92     &--- &--- &  0.998  \\
							\hline \hline
							\multicolumn{10}{c}{$n = 500$}\\
							\hline \\
							& $\hat{\alpha}$& $\hat{\beta}$ & $\hat{\theta}$ & $\hat{\alpha}_{MK}$ & $\hat{\beta}_{MK}$ & $\hat{\theta}_{MK}$ & $\tilde{\alpha}_{MAR}$ & $\tilde{\beta}_{MAR}$ & $\tilde{\theta}_{MAR}$ \\[1mm]		
							\hline\\ 	
							Bias & -0.001   &   -0.026  &0.003  & -0.042  &   0.041&  0.022 & -1.003 & 1.000 &0.146 \\
							Stdev  &  0.203  &   0.139  &  0.071 & 0.172   &  0.100   &  0.067   &  0.048 &  0.088 &  0.199  \\
							MSE  & 0.041 &  0.020  & 0.005  &  0.031  & 0.011   &  0.005 & 1.010 & 1.009 &0.061 \\
							CP  & --- & ---  &  0.944  & --- & ---  & 0.946   & ---&--- & 1.000  \\	
							\hline \hline  \multicolumn{10}{c}{}	
							\\[1mm]
							\hline \hline 
							\multicolumn{10}{c}{$n = 1000$}\\ 
							\hline	\\
							& $\hat{\alpha}$& $\hat{\beta}$ & $\hat{\theta}$ & $\hat{\alpha}_{MK}$ & $\hat{\beta}_{MK}$ & $\hat{\theta}_{MK}$ & $\tilde{\alpha}_{MAR}$ & $\tilde{\beta}_{MAR}$ & $\tilde{\theta}_{MAR}$ \\[1mm]		
							\hline\\ 		
							Bias & 0.010   &  -0.034   & -0.001  & -0.027 & 0.024 & 0.011 & -1.000 & 0.997 & 0.134 \\
							Stdev  &  0.262  &  0.264 & 0.052   &0.122  &0.070  & 0.048  &  0.035 &  0.065 & 0.148  \\
							MSE  &  0.068  & 0.070  &  0.002  & 0.015 & 0.005  &0.002  &  1.003& 1.000 & 0.039\\
							CP  & --- & ---  & 0.936    & --- & ---  &  0.932 & ---& ---& 1.000\\	
							\hline\hline
						\end{tabular}}
						{\fontsize{9.5pt}{14pt} \selectfont  Stdev: standard deviation ; MSE: mean squared error;
							CP: coverage probability. The bandwith used in computing the nonparametric kernel estimators $(\hat{\alpha}_{MK},\hat{\beta}_{MK},\hat{\theta}_{MK})$ is $h=0.2$.	
						}
					\end{threeparttable}
				\end{table}

				\section{Simulations}
				After establishing the large sample properties of the proposed estimator, we now evaluate its finite sample performance through a small scale simulation
				study. We consider four scenarios. In all scenarios, the parameter of
				interest is $\theta _{0}=\mathbb{E}[Y]$ and the sample size is set
				respectively at $N=200,500$ and $1000$.

				\begin{itemize}
					\item {\color{black} \textbf{Scenario I}: $X$ is generated from the normal distribution $N(0,1)$, and the outcome $Y$ is generated from the  normal distribution with mean $X+1$ and unit variance, i.e.  $Y\sim N(X+1,1)$. The relationship between the outcome variable and the covariate is linear, and the  distribution of outcome is normal. The missing mechanism is modeled by 
						$$\mathbb{P}(T=1|Y,X)=[1+\exp(\alpha_0 +\beta_0 Y)]^{-1}\ ,$$
						with the true value $(\alpha_0,\beta_0)=(0,-1.2)$. The true value of the parameter of interest is $\theta_0=\mathbb{E}[Y]=1$}.\\
					\item \textbf{Scenario II}:   $X$ is generated from the normal distribution $%
					N(0,1)$, and the outcome $Y$ is generated from the normal distribution with
					mean $X^{2}+1$ and unit variance, i.e. $Y\sim N(X^{2}+1,1)$. Thus the
					relationship between the outcome variable and the covariate is nonlinear,
					and the distribution of outcome is non-normal. The missing mechanism is
					modeled as   
					\begin{equation*}
					\mathbb{P}(T=1|Y,X)=[1+\exp (\alpha _{0}+\beta _{0}Y)]^{-1}\ 
					\end{equation*}%
					with the true value $(\alpha _{0},\beta _{0})=(1.25,-1.2)$. The
					true value of the parameter of interest is $\theta _{0}=\mathbb{E}[Y]=2$.  \\
					\item \textbf{Scenario III}.  The design follows \cite{qin2002estimation}. We
					generate the outcome from  
					\begin{equation*}
					Y=0.1X^{2}+ZX^{1/2}/5\ ,
					\end{equation*}%
					where $Z$ and $X$ are independent, $Z$ is standard normal random variable,
					and $X$ follows the $\chi _{(6)}^{2}/2$ distribution. The missing mechanism
					is modeled as  
					\begin{equation*}
					\mathbb{P}(T=1|Y,X)=\left[ 1+\exp (\alpha _{0}+\beta _{0}Y)\right] ^{-1}\ 
					\end{equation*}%
					with the true value $(\alpha _{0},\beta _{0})=(3,-1)$. The true value of the
					target parameter is $\theta _{0}=\mathbb{E}[Y]=1.2$.\\
					
					\item \textbf{Scenario IV}. The design is similar to that in \cite{kang2007demystifying}. $\boldsymbol{Z}=(Z_{1},Z_{2})$ is generated from the standard
					bivariate normal distribution, and $Y$ is generated from the normal
					distribution with mean $2+Z_{1}$ and unit variance. The missing mechanism is
					modeled as  
					\begin{equation*}
					\mathbb{P}(T=1|Y,X_{1},X_{2})=[1+\exp (\alpha _{0}Z_{1}+\beta _{0}Y)]^{-1}\ 
					\end{equation*}%
					with $(\alpha _{0},\beta _{0})=(1,-1)$. The true value of the parameter of
					interest is $\theta _{0}=\mathbb{E}[Y]=2$. Instead of directly observing
					covariates $\boldsymbol{Z}$, we observe a non-linear transformation of $%
					\boldsymbol{Z}$: $X_{1}=\exp (Z_{1}/2)$ and $X_{2}=Z_{2}/(1+\exp (Z_{1}))$.
				\end{itemize} \
				In all scenarios, we generate $J=500$ random samples, and for each sample, we compute the following three estimators:
				\begin{enumerate}
					\item Naive estimator. We compute the missing at random estimator $(%
					\tilde{\alpha}_{MAR},\tilde{\beta}_{MAR},\tilde{\theta}_{MAR})$ as   
					\begin{equation*}
					\tilde{\theta}_{MAR}=\frac{1}{N}\sum_{i=1}^{N}\frac{T_{i}}{\pi (\boldsymbol{X}%
						_{i};\tilde{\alpha}_{MAR},\tilde{\beta}_{MAR})}Y_{i}\ ,
					\end{equation*}%
					where $\pi (\boldsymbol{X}_{i};\tilde{\alpha}_{MAR},\tilde{\beta}_{MAR})$ is
					an estimated response model. In Scenarios I, II \& III, $%
					\pi (\boldsymbol{X}_{i};\tilde{\alpha}_{MAR},\tilde{\beta}_{MAR})=\left[
					1+\exp (\tilde{\alpha}_{MAR}+\tilde{\beta}_{MAR}X_{i})\right] ^{-1}$ and 
					in Scenario IV $\pi (\boldsymbol{X}_{i};\tilde{\alpha}_{MAR},\tilde{\beta}%
					_{MAR})=\left[ 1+\exp (\tilde{\alpha}_{MAR}Z_{1i}+\tilde{\beta}_{MAR}X_{2i})%
					\right] ^{-1}$, where $(\tilde{\alpha}_{MAR},\tilde{\beta}_{MAR})$ are
					estimated by GMM. 
					\item MK2 estimator. We compute $(\hat{\alpha}_{MK},\hat{\beta}_{MK},\hat{%
						\theta}_{MK})$ using the approach of \cite{morikawa2016semiparametric}, i.e.	 $(\hat{\alpha}_{MK},\hat{\beta}_{MK},\hat{%
						\theta}_{MK})$ is the solution of $$\sum_{i=1}^N\left(\hat{\boldsymbol{S}}_{1}(T_i,\boldsymbol{Z}_i;\alpha,\beta)^{\top}, \hat{S}_2(T_i,\boldsymbol{Z}_i;\alpha,\beta,\theta)\right)^{\top}=0\ ,$$ where
					\begin{align*}
					&\hat{\boldsymbol{S}}_{1}(T,\boldsymbol{Z};\alpha,\beta)=-\left(1-\frac{T}{\pi(\boldsymbol{Z};\alpha,\beta)}\right)\mathbb{E}^\star\left[\frac{\nabla_{\gamma}\pi(\boldsymbol{Z};\alpha,\beta)}{1-\pi(\boldsymbol{Z};\alpha,\beta)}\bigg|\boldsymbol{X}\right]\ , \\
					&\hat{S}_2(T,\boldsymbol{Z};\alpha,\beta,\theta)=-\frac{T}{\pi(\boldsymbol{Z};\alpha,\beta)}U(\boldsymbol{Z})+\theta-\left(1-\frac{T}{\pi(\boldsymbol{Z};\alpha,\beta)}\right) \mathbb{E}^\star\left[U(\boldsymbol{Z})|\boldsymbol{X}\right]\ ,
					\end{align*}
					and for any function $g(\boldsymbol{Z})$ the quantity $\mathbb{E}^\star[g(\boldsymbol{Z})|\boldsymbol{X}]$ is defined by	
					\begin{align*}
					&\mathbb{E}^\star[g(\boldsymbol{Z})|\boldsymbol{X}=\boldsymbol{x}]:=\frac{\sum_{j=1}^NT_jK_h(\boldsymbol{x}-\boldsymbol{X}_j)T_j\pi(\boldsymbol{Z}_j;\alpha,\beta)^{-1}O(\boldsymbol{x},Y_j;\alpha,\beta)g(\boldsymbol{x},Y_j)}{\sum_{j=1}^NK_h(\boldsymbol{x}-\boldsymbol{X}_j)T_j\pi(\boldsymbol{Z}_j;\alpha,\beta)^{-1}O(\boldsymbol{x},Y_j;\alpha,\beta)}\  ; \\
					&O(\boldsymbol{z};\alpha,\beta)=\frac{1-\pi(\boldsymbol{z};\alpha,\beta)}{\pi(\boldsymbol{z};\alpha,\beta)}  \ ,
					\end{align*}
					$K_h(\boldsymbol{x}-\boldsymbol{w})=K\left((\boldsymbol{x}-\boldsymbol{w}/h)\right)$, $K(\cdot)$ is Gaussian kernel function and $h$ is the bandwidth.
					\item Our GMM estimator. We compute $(\hat{\alpha},\hat{\beta},\hat{\theta})$
					using the proposed approach and the covariate balancing approach to select $K
					$, {\color{black}with $\bar{K}=7$ in Scenarios I, II, III, and with $\bar{K}=10$ in Scenario IV. Here $\bar{K}$ is the maximal number of candidate moments to be considered.} 
				\end{enumerate}
				The simulation results (the bias, the standard deviation (Stdev), the mean squared error (MSE), and the coverage probability (CP) (for significance level $\alpha =0.05$) of the point estimates) for all scenarios are reported in Tables 1, 2, 3, and 4 respectively. {\color{black} The histogram of selected $K'$s  (based on $500$ Monte Carlo samples) in all scenarios is reported in Figure 1.} Glancing at these tables, we find: 
				\begin{enumerate}
					\item In all scenarios, the naive estimator using the missing at random assumption has a large bias, because this assumption does not hold.
					
					\item In all scenarios, our proposed estimator of $\mathbb{E}[Y]$
					out-performs the MK estimator. 
					\item In all scenarios, our proposed variance estimator has coverage probability close to $95\%$, even the sample size is small. The MK's variance estimator performs well in Scenario IV, but badly in other scenarios: in Scenarios I, the coverage probability using MK's approach   converges to $90\%$ rather than $95\%$; in Scenarios II, the CP values are far from $95\%$ in Scenario 2 no matter the sample size is small or large; in Scenarios III, the MK's variance estimaotr is consistent only when  the sample size is large.
					\item {\color{black}When the sample size  is small the optimal $K$ tends to be $2$. When the sample size is large, the optimal $K$ tends to be $3$. The growing rate  of $K$ is extremely slow comparing to that of the sample size $n$, which is consistent with our theoretical Assumption \ref{as:K&N}.} 
				\end{enumerate}
				These results clearly show that the proposed approach has better finite sample performance.

				\begin{figure}[!htp]
					\caption{Histogram of $K$}
					\begin{center}
						\includegraphics[height=21cm,width=15cm, scale=1.0]{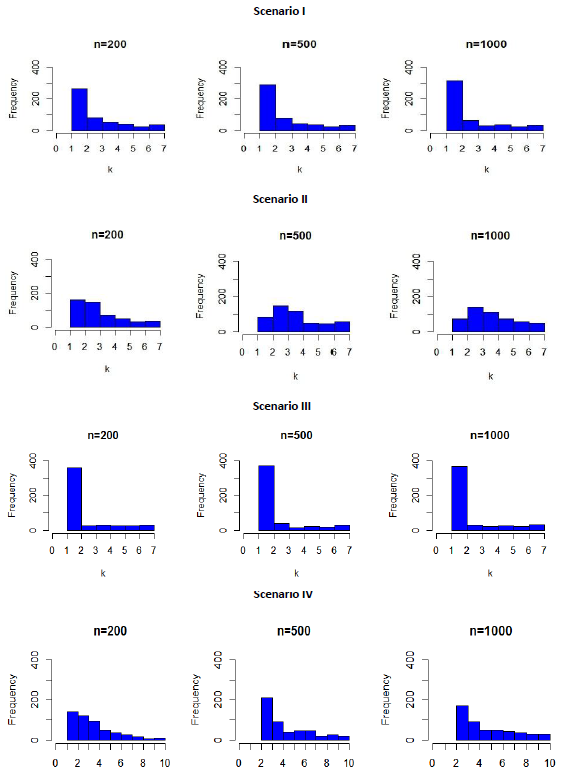}
						\footnotesize \\[4mm]  The Monte Carlo sample size used to  plot the histogram of $K$ is $J=500$. 
					\end{center}
				\end{figure}

				\newpage	
				\section{Discussion}
				The data missing not at random problem is common in applications. \cite
				{morikawa2016semiparametric} studies the efficient estimation of a class of missing not at random problems. But their approach requires nonparametric estimation of the conditional density function and thus suffers from the curse of dimensionality and smoothing parameter selection problem. In this
				paper, we study the same class of missing not at random problems but present a much simpler and more natural efficient estimator. Our approach is based
				on a parametric moment restriction model that does not require nonparametric estimation and hence does not suffer from the curse of dimensionality problem nor the bandwidth selection problem. Indeed the simulation results confirm that the proposed approach out-performs their approach in finite samples. The GMM approach is also easy to adapt to stratified sampling and other sampling schemes common in survey data. \\ 
				
				Both approaches require correct parameterization of the propensity score function. If the propensity score function is misspecified, then both approaches yield inconsistent estimates. There is some attempt in the literature to mitigate this problem. For instance, Zhao and Shao (2015) introduce a partial linear index to model missing mechanism. The proposed approach can be extended in this direction. Such extension shall be pursued in a future study.

				\section{Appendix}
				\subsection{Assumptions}
				We first introduce the smoothness classes of functions used in the nonparametric estimation; see e.g. \cite{stone1982optimal, stone1994use}, \cite{robinson1988root}, \cite{newey1997convergence}, \cite{horowitz2012semiparametric} and \cite{chen2007large}. Suppose that $\mathcal{X}$ is the Cartesian product of $r$-compact intervals. Let $0<\delta \leq 1$. A fucntion $f$ on $\mathcal{X}$ is said to satisfy a H$\ddot{\text{o}}$lder condition with exponent $\delta$ if there is a positive constant $L$ usch that $\|f(\boldsymbol{x}_1)-f(\boldsymbol{x}_2)\|\leq L \|\boldsymbol{x}_1-\boldsymbol{x}_2\|^{\delta}$ for all $\boldsymbol{x}_1,\boldsymbol{x}_2\in\mathcal{X}$. Given  a $r$-tuple $\boldsymbol{\alpha}=(\alpha_1,...,\alpha_r)$ of nonnegative integer, denote $[\boldsymbol{\alpha}]=\alpha_1+\cdots+\alpha_r$ and let $D^{\boldsymbol{\alpha}}$ denote the differential operator defined by $D^{\boldsymbol{\alpha}}=\frac{\partial^{[\boldsymbol{\alpha}]}}{\partial x_1^{\alpha_1}\cdots \partial x_r^{\alpha_r}}$, where $\boldsymbol{x}=( x_1,..., x_r)$. 
				\begin{definition}
					\emph{Let $s$ be a nonnegative integer and $s:=s_0+\delta$.  The function $f$ on $\mathcal{X}$ is said to be} $s$-smooth \emph{if it is $s$ times continuously differentiable on $\mathcal{X}$ and $D^{\boldsymbol{\alpha}}f$ satisfies a  H$\ddot{\text{o}}$lder condition with exponent $\delta$ for all $\boldsymbol{\alpha}$ with $[\boldsymbol{\alpha}]=s_0$}. 
				\end{definition}

				The following notations are needed for presenting the efficiency bounds:
				\begin{align} & O (\boldsymbol{Z}):=\frac{1 -\pi  (\boldsymbol{Z} ;\gamma _{0})}{\pi  (\boldsymbol{Z} ;\gamma _{0})} ,\text{\quad }\boldsymbol{S}_{0} (\boldsymbol{Z}):= -\frac{ \nabla _{\gamma }\pi  (\boldsymbol{Z} ;\gamma _{0})}{1 -\pi  (\boldsymbol{Z} ;\gamma _{0})}  \label{def:os}\ , \\
				&m(\boldsymbol{X}):=\frac{\mathbb{E}[O(\boldsymbol{Z})\boldsymbol{S}_{0} (\boldsymbol{Z})\vert \boldsymbol{X}]}{\mathbb{E}[O(\boldsymbol{Z})\vert \boldsymbol{X}]} ,\text{\quad }R (\boldsymbol{X}):=\frac{\mathbb{E}[O(\boldsymbol{Z})U (\boldsymbol{Z})\vert \boldsymbol{X}]}{\mathbb{E}[O(\boldsymbol{Z})\vert \boldsymbol{X}]}\;  \label{def:osmr}\ , \\
				& \boldsymbol{S}_{1} (T ,\boldsymbol{Z} ;\gamma _{0}) :=\left (1 -\frac{T}{\pi  (\boldsymbol{Z} ;\gamma _{0})}\right ) m (\boldsymbol{X})\;\text{,} \label{def:S1}\\
				& S_{2} (T ,\boldsymbol{Z} ;\gamma _{0} ,\theta _{0}) := -\frac{T}{\pi  (\boldsymbol{Z} ;\gamma _{0})} U (\boldsymbol{Z}) +\theta _{0} -\left (1 -\frac{T}{\pi  (\boldsymbol{Z} ;\gamma _{0})}\right ) R (\boldsymbol{X})\;\text{.} \label{def:S2}
				\end{align}
				
				The following assumptions are required in this paper:
				\begin{assumption}
					\label{as:id} There exists a nonresponse instrumental variable $\mathbf{X}_{\mathbf{2}}$, i.e., $\mathbf{X} =(\mathbf{X}_{1}^{^{ \intercal }} ,\mathbf{X}_{2}^{^{ \intercal }})^{^{ \intercal }}\text{,}$ such that $\boldsymbol{X}_{2}$ is independent of $T$ given $\boldsymbol{X}_{1}$ and $Y$; furthermore, $\boldsymbol{X}_{2}$ is correlated with $Y$. 
				\end{assumption}

				\begin{assumption}
					\label{as:suppX} The support of $\mathbf{X}$, which is denoted by $\mathcal{X}$, is a Cartesian product of $r$-compact intervals, and we denote $\boldsymbol{X}=(X_1,...,X_r)^{\top}$.
				\end{assumption}
				
				\begin{assumption}
					\label{as:proj_smooth} The functions $\mathbb{E}[O(\boldsymbol{Z})S_{0} (\boldsymbol{Z})\vert 
					\bold {X}
					=\boldsymbol{x}]$, $\mathbb{E}[O(\boldsymbol{Z})U (\boldsymbol{Z})\vert 
					\bold {X}
					=\boldsymbol{x}]$ and $\mathbb{E}[O(\boldsymbol{Z})\vert \boldsymbol{X} =\boldsymbol{x}]$ are $s$-smooth in  $\boldsymbol{x}$, where $s > 0$. 
				\end{assumption}
				
				\begin{assumption}\label{as:eigen}
					There exists two finite positive constants $\underline{a}$ and $\overline{a}$ such that the smallest (resp. largest) eigenvalue of $\mathbb{E}[u_K(\boldsymbol{X})u_K^{\top}(\boldsymbol{X})]$ is bounded away from $\underline{a}$ (resp.  $\overline{a}$) uniformly in $K$, i.e.,  $$0<\underline{a}\leq \lambda_{\min}(\mathbb{E}[u_{K}(\boldsymbol{X})u_{K} (\boldsymbol{X})^{\top }])\leq \lambda_{\max}(\mathbb{E}[u_{K}(\boldsymbol{X})u_{K} (\boldsymbol{X})^{\top }])\leq \overline{a}<\infty\ .$$ 
				\end{assumption}
				\textbf{Remark}: Asssumption \ref{as:eigen} implies that following results:
				\begin{enumerate}
					\item 	\begin{align}\label{bound:Eu_K^2}
					\mathbb{E}[\|u_K(\boldsymbol{X})\|^2]=\tr\left(\mathbb{E}\left[u_K(\boldsymbol{X})u_K(\boldsymbol{X})^{\top}\right]\right)=O(K)\ ;
					\end{align}
					\item the matrices
					$\bar{a}\cdot I_{K\times K}-\mathbb{E}[u_{K}(\boldsymbol{X})u_{K} (\boldsymbol{X})^{\top }]$ and  $\mathbb{E}[u_{K}(\boldsymbol{X})u_{K} (\boldsymbol{X})^{\top }]-\underline{a}\cdot I_{K\times K}$ are positive definite, and
					\begin{align}\label{bound_u_kK}
					\underline{a}\leq 	\inf_{k\in\{1,...,K\}}\mathbb{E}[u_{kK}(\boldsymbol{X})^2]\leq 	\sup_{k\in\{1,...,K\}}\mathbb{E}[u_{kK}(\boldsymbol{X})^2] \leq \overline{a}\ .
					\end{align}
				\end{enumerate}

				\begin{assumption}
					\label{as:iid}The full data $\{(T_{i} ,\boldsymbol{X}_{i} ,Y_{i})\}_{i =1}^{N}$ are independently and identically distributed. 
				\end{assumption}

				\begin{assumption}
					\label{as:boundS}$	\boldsymbol{S}_{eff}(T,\boldsymbol{Z};\gamma,\theta%
					):=(\boldsymbol{S}_{1}^{^{\intercal}}(T,\boldsymbol{Z};\gamma
					),S_{2}(T,\boldsymbol{Z};\gamma,\theta))^{^{\intercal}}$ 
					is continuously differentiable
					at each $(\gamma  ,\theta ) \in \Gamma  \times \Theta $ with probability one, and $\mathbb{E} \left [ \partial \boldsymbol{S}_{e f f} (\gamma  ,\theta )/ \partial (\gamma ^{\top } ,\theta )\right ]$ is nonsingular at $(\gamma _{0} ,\theta _{0})$. 
				\end{assumption}

				\begin{assumption}
					\label{as:pi} The response probability $\pi $ satisfies the following conditions:
					\begin{enumerate}
						\item there exists two positive constants $\bar{c}$ and $\underline{c}$ such that $0 <\underline{c} \leq \pi  (\boldsymbol{x} ,y ;\gamma ) \leq \bar{c} <1$ for all $\gamma  \in \Gamma $ and $(\boldsymbol{x} ,y) \in \mathcal{X} \times \mathbb{R}$;  
						
						\item the propensity score $\pi  (\boldsymbol{x},y;\gamma )$ is twice continuously differentiable in $\gamma  \in \Gamma $, and the derivatives are uniformly bounded. 
						\item for any $\gamma\in \Gamma$, the conditional functions $\mathbb{E}\left[1-\frac{T}{\pi(\boldsymbol{Z};\gamma)}|\boldsymbol{X}=\boldsymbol{x}\right]$ and $\mathbb{E}\left[\frac{\nabla_{\gamma}\pi(\boldsymbol{Z};\gamma)}{\pi(\boldsymbol{Z};\gamma)}\bigg|\boldsymbol{X}=\boldsymbol{x}\right]$ are $s$-smooth  in $\boldsymbol{x}$, where $s> 0$. 
					\end{enumerate}
				\end{assumption}
				\begin{assumption}
					\label{as:K&N} Suppose $K \rightarrow \infty $ and 	{\color{black}$K^3/N \rightarrow 0$}. 
				\end{assumption}
				The Assumption \ref{as:id} is used for the identification
				of the model, which was discussed in \cite{wang2014instrumental}. Assumptions
				\ref{as:suppX} and \ref{as:proj_smooth} are required for uniform boundedness
				of approximations.   Assumption \ref{as:eigen} is a standard assumption used in nonparametric sieve approximation, see also  \cite{newey1997convergence}. Assumption \ref{as:iid} is a standard condition for statistical sampling. Assumptions \ref{as:boundS}-\ref{as:pi}
				are required for the convergence of our estimator as well as the boundness of the asymptotic variance. Assumption \ref{as:K&N}
				is the same as Assumption 2 in \cite{newey2003instrumental}, it is required
				for controlling the stochastic order of the residual terms, which is desirable in practice because $K$ grows very slowly with $N$ so a relatively small number of moment conditions is sufficient for the method proposed to perform well. 
				
				{\color{black}\subsection{Discussion on $u_K$}\label{sec:uK}	
					To construct the GMM estimator, we need to specify  the matching function $u_K(\boldsymbol{X})$. Although the approximation theory is derived for general sequences of approximating functions, the most common class of functions
					are power series. Suppose the dimension of covariate $\boldsymbol{X}$ is $r\in\mathbb{N}$, namely $\boldsymbol{X}=(X_1,...,X_r)^{\top}$.  Let $\lambda  =(\lambda _{1} ,\ldots  \lambda _{r})^{\top }$ be an $r$-dimensional vector of nonnegative integers (multi-indices), with norm $\vert \lambda \vert  =\sum _{j =1}^{r}\lambda _{j}$. Let $(\lambda  (k))_{k =1}^{\infty }$ be a sequence that includes all distinct multi-indicesand satisfies $\vert \lambda  (k)\vert  \leq \vert \lambda  (k +1)\vert $, and let $\boldsymbol{X}^{\lambda } =\prod _{j =1}^{r}X_{j}^{\lambda _{j}}$. For a sequence $\lambda  (k)$ we consider the series $u_{k K} (\boldsymbol{X}) =\boldsymbol{X}^{\lambda  (k)}$, $k\in\{1,...,K\}$.  \cite{newey1997convergence}
					showed the following property for the power series:  there exists a universal constant $C >0$ such that
					\begin{align}\zeta  (K):=\underset{\boldsymbol{x} \in \mathcal{X}}{\sup }\Vert u_{K} (\boldsymbol{x})\Vert  \leq C K \label{eq:zeta}\ , \end{align}where $\Vert  \cdot \Vert $ denotes the usual matrix norm $\Vert A\Vert  =\sqrt{\tr (A^{\top } A)}$. \\
					
					Another important issue is choosing  the number of the matching function $K$ in finite sample experiment. \cite{donald2009choosing} proposed a strategy for an appropriate choice of $K$ by minimizing the higher order MSE defined in \eqref{eq:criteria_K}, and the following notations are needed to  describe this criteria:
					\begin{align*} & \rho  (T_{i} ,\boldsymbol{X}_{i} ,Y_{i} ;\check{\gamma }) =1 -\frac{T_{i}}{\pi  (\boldsymbol{X}_{i} ,Y_{i} ;\check{\gamma })} ,\text{\quad }\hat{\Upsilon }_{K \times K} =\frac{1}{N} \sum _{i =1}^{N}\rho  (T_{i} ,\boldsymbol{X}_{i} ,Y_{i} ;\check{\gamma })^{2} u_{K} (\boldsymbol{X}_{i})^{ \otimes 2}\text{,} \\
					& \widehat{\Gamma }_{K \times p} =\frac{1}{N} \sum _{i =1}^{N}u_{K} (\boldsymbol{X}_{i})  \nabla _{\gamma }\rho  (T_{i} ,\boldsymbol{X}_{i} ,Y_{i} ;\check{\gamma })^{\top } ,\text{\quad }\hat{\Omega }_{p \times p} =(\hat{\Gamma }_{K \times p})^{\top } \hat{\Upsilon }_{K \times K}^{ -1} \hat{\Gamma }_{K \times p}\text{,} \\
					& \tilde{\mathbf{d}}_{i} =(\hat{\Gamma }_{K \times p})^{\top } \left (\frac{1}{N} \sum _{j =1}^{N}u_{K} (\boldsymbol{X}_{j})^{ \otimes 2}\right )^{ -1} u_{K} (\boldsymbol{X}_{i}) ,\text{\quad }\tilde{\mathbf{\eta }}_{i} = \nabla _{\gamma }\rho  (T_{i} ,\boldsymbol{X}_{i} ,Y_{i} ;\check{\gamma }) -\tilde{\mathbf{d}}_{i}\text{,} 
					\end{align*}
					\begin{align*}
					& \hat{\xi }_{i j} =\frac{1}{N} u_{K} (\boldsymbol{X}_{i})^{\top } \hat{\Upsilon }_{K \times K}^{-1} u_{K} (\boldsymbol{X}_{j}) ,\text{\quad }\widehat{
						\bold {D}
					}_{i}^{ \ast } =(\hat{\Gamma }_{K \times p})^{\top } \hat{\Upsilon }_{K \times K}^{ -1} u_{K} (\boldsymbol{X}_{i})\;\text{.}\end{align*}
					
					\subsection{Semiparametric Efficiency Bounds}
					The following lemma is Theorem 1 in \cite{morikawa2016semiparametric}.
					\begin{lemma}[\cite{morikawa2016semiparametric}]
						\label{lemma:MK} The  efficient variance bounds of $(\gamma _{0} ,\theta _{0})$ is	$\boldsymbol{V}_{eff}:=\mathbb{E}\left[\boldsymbol{S}_{eff}(T,\bold{Z};\gamma_0,\theta_0)^{\otimes 2}\right]^{-1}$,  where $\boldsymbol{S}_{eff}=(\boldsymbol{S}_1^{\top},S_2)^{\top}$ and $\boldsymbol{S}_1$, $S_2$ are defined in \eqref{def:S1} and \eqref{def:S2} respectively.
					\end{lemma} 
					
					Let $\boldsymbol{V}_{\gamma _{0}}$ (resp. $V_{\theta _{0}}$) be the efficient variance bound of  $\gamma_0$ (resp. $\theta_0$). After some simple computation, we can find out 
					\begin{align}\boldsymbol{V}_{\gamma _{0}} = & \mathbb{E} \left [\frac{1 -\pi  (\boldsymbol{Z} ;\gamma _{0})}{\pi  (\boldsymbol{Z} ;\gamma _{0})} m (\boldsymbol{X})^{ \otimes 2}\right ]^{ -1} \label{effbound_gamma}\end{align}
					and
					\begin{align}V_{\theta _{0}} =V a r \left (S_{2} (T ,\boldsymbol{Z} ;\gamma _{0} ,\theta _{0}) -\mathbf{\kappa }^{^{ \intercal }} \boldsymbol{S}_{1} (T ,\boldsymbol{Z} ;\gamma _{0})\right )\;\text{.} \label{effbound_theta}\end{align}
					where
					\begin{align}\mathbf{\kappa }^{^{ \intercal }}:=\mathbb{E} \left [\frac{ \nabla _{\gamma }\pi  (\boldsymbol{Z} ;\gamma _{0})^{^{ \intercal }}}{\pi  (\boldsymbol{Z} ;\gamma _{0})} \{R (\boldsymbol{Z}) -U (\boldsymbol{X})\}\right ] \cdot \mathbb{E} \left [\frac{m (\boldsymbol{X})}{\pi  (\boldsymbol{Z} ;\gamma _{0})}  \nabla _{\gamma }\pi  (\boldsymbol{Z} ;\gamma _{0})^{^{ \intercal }}\right ]^{ -1}\;\text{.} \label{def:kappa}\end{align}

\bibliographystyle{dcu}
\bibliography{Semiparametric}

\end{document}